\documentstyle[12pt]{article} \textwidth=6in \oddsidemargin=0in
\begin{document} 
\newcommand{\dl}{\delta} \newcommand{\iy}{\infty} \newcommand{\La}{\Lambda} 
\newcommand{\pl}{\partial} \newcommand{\noi}{\noindent} \renewcommand{\sp}{\vskip2ex}
\newcommand{\bq}{\begin{equation}} \newcommand{\eq}{\end{equation}}
\newcommand{\ra}{\rightarrow} \newcommand{\al}{\alpha} \newcommand{\th}{\theta}
\newcommand{\bR}{{\bf R}} \newcommand{\bZ}{{\bf Z}} \newcommand{\sr}{^{(r)}}
\newcommand{\bC}{{\bf C}}\newcommand{\bD}{{\bf D}}\newcommand{\bT}{{\bf T}}
\newcommand{\nm}{\parallel} \newcommand{\la}{\lambda} \newcommand{\ph}{\varphi}
\newcommand{\om}{\omega} \newcommand{\inv}{^{-1}}\newcommand{\ve}{\varepsilon}
\newcommand{\ov}{\over} \newcommand{\Om}{\Omega}\newcommand{\be}{\beta}
\newcommand{\ep}{\varepsilon} \renewcommand{\nu}{\sigma}
\newcommand{\ch}{\raisebox{.4ex}{$\chi$}} \newcommand{\tl}{\widetilde}
\begin{center}{\large \bf Some Classes of Solutions to the Toda Lattice Hierarchy}
\end{center}\vskip4ex

\begin{center}{{\bf Harold Widom}\\
{\it Department of Mathematics\\
University of California\\ Santa Cruz, CA 95064, USA\\
e-mail address: widom@math.ucsc.edu}}\end{center}\sp
\begin{abstract}
We apply an analogue of the Zakharov-Shabat dressing method to obtain infinite matrix
solutions to the Toda lattice hierarchy. Using an operator transformation 
we convert some of these into solutions in terms
of integral operators and Fredholm determinants. Others are converted into 
a class of operator solutions to the $l$-periodic Toda hierarchy. 

\end{abstract}
\renewcommand{\theequation}{0.\arabic{equation}}
\noi{\bf 0. Introduction}\sp

We begin by recalling some terminology: The shift matrix $(\dl_{i+1,j})$ is
denoted by $\La$. (All matrices are doubly-infinite unless otherwise stated.) Any
matrix $A$ has a representation as a formal sum $\sum_{i=-\iy}^{\iy}a_i\La^i$ where the 
$a_i$ are diagonal matrices. Two of these may be multiplied if for both matrices the 
indices corresponding to nonzero components are bounded above (or below), or if for 
one of the matrices these indices are bounded above and below. If $A$ is triangular 
then it is invertible if and only if each daiagonal entry of $a_0$ is nonzero. The 
upper-triangular and strictly lower-triangular projections of $A$ are defined by
\[A_+=\sum_{i=0}^{\iy}a_i\La^i,\quad A_-=\sum_{i=-\iy}^{-1}a_i\La^i.\]

A solution to the Toda lattice (TL) hierarchy \cite{UT} is a family of matrices of the 
form
\bq B_n=\sum_{i=0}^{n-1}b_{i,n}\La^i+\La^n,\quad C_n=\sum_{i=-n}^{-1}c_{i,n}\La^i
\label{BCform}\eq
which satisfy the 2-dimensional TL equations

\[\pl_{x_n}B_m-\pl_{x_m}B_n+[B_m,B_n]=0,\]
\[\pl_{y_n}C_m-\pl_{y_m}C_n+[C_m,C_n]=0,\]
\[\pl_{y_n}B_m-\pl_{x_m}C_n+[B_m,C_n]=0.\]

In the first section we apply a discrete version of the dressing method of Zakharov and 
Shabat \cite{ZS} (see also the nice exposition in \cite{PS}) to obtain infinite
matrix solutions to the TL hierarchy. By this we mean that the entries of the $B_n$ and 
$C_n$ are themselves expressed in terms of infinite matrices, analogous to the operators 
whose Fredholm determinants give solutions to the KP hierarchy.

Matrix solutions to the semi-infinite TL hierarchy were obtained in \cite{P1} and \cite{AV}.
The methods here and in these references are different although the latter also begins with
the factorization of a type of moment matrix. 

In the following section we obtain operator solutions to the TL hierarchy, in which the
entries of $B_n$ and $C_n$ are expressible in terms of integral operators and the diagonal
entries of the $B_n$ are given in terms of Fredholm determinants. These are obtained
from the matrix solutions by applying the fact that the inverses of the operators
$I-AB$ and $I-BA$ may be expressed in terms of each other, and that they generally
have the same determinant. The simplest integral operators which arise have kernel
\[{e^{\sum (x_n-y_n)(u^n-u^{-n}+v^n-v^{-n})/2}\over 1-uv}\]
and act on the space $L_2(\nu)$ where $\nu$ is a measure supported in the unit disc 
of the complex plane. When $\nu$ equals a function
$p(u)^2$ times Lebesgue measure we obtain an equivalent operator which acts on $L_2$
of Lebesgue measure when the kernel given above is multiplied by $p(u)\,p(v)$. If we set
$t=x_1-y_1$ (ignoring the other parameters) these operators give Fredholm determinant
solutions to the Toda equations
\[{d^2 q_k\over d t^2}=e^{q_{k-1}-q_k}-e^{q_k-q_{k+1}}\qquad(k\in\bZ).\]
When $\nu$ is a discrete measure supported on $N$ points the Fredholm determinants are
finite determinants and these give the familiar $N$-soliton solutions. 

In the next section we use the same device to obtain another class of operator solutions 
which includes solutions to the $l$-periodic Toda hierarchy. The simplest operators here
have kernel 
\[{e^{\sum [x_n(1-\om^{-n})(u^n+v^n)+y_n(1-\om^n)(u^{-n}+v^{-n})]/2}\over u-\om v},\]
where $\om$ is an $l$th root of unity, and act on $L_2(\nu)$ where now $\nu$ is a 
measure on $\bR^+$. Again there
is the special case where the operator acts on the usual $L_2(\bR^+)$ space and the
kernel is multiplied by $p(u)\,p(v)$.  When $\nu$ is a discrete 
measure supported on finitely many points the solutions are again expressed in terms of 
finite determinants. A related class of kernels gives solutions to the 
cylindrical Toda equations
\[{d^2 q_k\over dt^2}+t\inv\,{dq_k\over dt}=e^{q_k-q_{k-1}}-e^{q_{k+1}-q_k}
\qquad(k\in\bZ).\]

Integral operator solutions to some analogous equations, or special cases, have also
appeared in the literature, for example \cite{BL,N,P,TW}. The methods here are
quite different. We mention that in \cite{TW} it 
was shown that the Fredholm determinants of the case $l=2$ of the last kernels 
gave solutions of the mKdV/sinh-Gordon hierarchies. In \cite{K} it was observed that 
those Fredholm determinant solutions were limits of $N$-soliton solutions. 
They are both special cases of the more general situation where the operator acts on an 
$L_2(\nu)$ space.

We shall see that it is easy {\em formally} to derive the solutions
to the TL hierarchy. But some justification is required, and the case of
the periodic TL hierarchy is a little tricky.\sp

\setcounter{equation}{0}\renewcommand{\theequation}{1.\arabic{equation}}

\noi{\bf I. Matrix solutions to the Toda hierarchy}\sp

In our discrete version of the dressing method we begin with a doubly-infinite matrix 
$F(i,j)$ for which there is a factorization 
$(I-F)=K_+^{-1}K_-$ with $K_+$ upper-triangular and $K_-$ of the form 
$I+$ strictly lower-triangular. More precisely, we require the relations
\bq K_+(I-F)=K_-,\quad (I-F)K_-\inv=K_+\inv. \label{FKK}\eq
(There is a difference, since not all matrix products are defined.)
To give conditions assuring the existence of such matrices $K_{\pm}$
we denote by $F_k$ the infinite matrix $F(i,j)$ with 
$i,j\geq k$ and denote by $\bZ_k$ the set of integers $\geq k$.\sp

\noi{\bf Lemma}. Assume that $F_k$ represents a bounded operator on $l_2(\bZ_k)$ for 
each $k$ and that each $I-F_k$ is invertible. Then there are triangular matrices $K_{\pm}$
of the form described such that all rows of $K_+$ and all columns of $K_-\inv$ belong to 
$l_2(\bZ)$ and such that the relations (\ref{FKK}) hold.\sp

\noi{\bf Proof}. We shall see what the matrices $K_{\pm}$ should be so that
(\ref{FKK}) holds, the actual verification then being a simple matter. 

If $K_+$ is upper-triangular then $K_+(I-F)$ is of the form $I+$ strictly lower-triangular
precisely when 
\[ K_+(k,j)-\sum_{i=k}^{\iy}K_+(k,i)F(i,j)=\dl_{j,k}\quad (j\geq k).\]
So we define $K_+$ to be the upper-triangular matrix with entries
\bq K_+(k,j)=(I-F_k)^{-1}(k,j)\quad (j\geq k).\label{Kkj}\eq
Note that each row of $K_+$ belongs to $l_2(\bZ)$. Note also that $K_+(k,k)$, the 
upper left-hand corner of $(I-F_k)\inv$, must be nonzero since the corresponding minor, 
the matrix $I-F_{k+1}$, is invertible.

Since $(I-F)K_-\inv$ should be upper triangular with $k,k$ entry
$K_+(k,k)\inv$ the entries of $K_-\inv$ must satisfy 
\[K_-\inv(i,k)-\sum_{j=k}^{\iy}F(i,j)K_-\inv(j,k)=K_+(k,k)\inv\,\dl_{i,k}\quad (i\geq k).\]
So we define $K_-\inv$ to be the lower-triangular matrix with entries
\[K_-\inv(i,k)=K_+(k,k)\inv\,\times\,i\mbox{th component of}\ (I-F_k)\inv\ve_k\quad 
(i\geq k),\]
where $\ve_k$ is the vector $(\dl_{i,k})$ of $\bZ_k$. Note that each column of 
$K_-\inv$ belongs to $l_2(\bZ)$.

The matrices $K_+$ and $K_-\inv$ defined above have the right form and their rows and 
columns, respectively, belong to $l_2(\bZ)$. Since each $F_k$ represents a bounded 
operator on $l_2(\bZ_k)$ the two matrix products $[K_+(I-F)]K_-\inv$ and $K_+[(I-F)K_-\inv]$
are well-defined and equal. It follows from our definitions that the first product
is of the form $I+$\nolinebreak strictly lower-triangular while the second is of the form
$I+$ strictly upper-triangular. Hence both products equal $I$, and this is equivelent
to the two relations (\ref{FKK}).\sp 

\noi {\bf Theorem 1}. Suppose, in addition to the hypotheses of the lemma, that 
\bq{\pl F\over\pl x_n} =F(i+n,j)-F(i,j-n),\quad {\pl F\over\pl y_n}=
F(i-n,j)-F(i,j+n).\label{F}\eq
Then the matrices $B_n$ and $C_n$ defined by 
\bq \pl_{x_n}-B_n=K_+(\pl_{x_n}-\La^n)K_+\inv,\quad  \pl_{y_n}-C_n=
K_+(\pl_{y_n}-\La^{-n})K_+\inv\label{BC+}\eq
are solutions to the TL hierarchy. Moreover, if we define
\[L=K_-\La K_-^{-1},\quad M=K_+\La^{-1} K_+^{-1},\]
then
\bq B_n=(L^n)_{+},\quad C_n=(M^n)_{-}.\label{BCLM}\eq\sp

\noi {\bf Proof}. The identities (\ref{F}) are equivalent to the statement that $F$ 
commutes with the operators $\pl_{x_n}-\La^n$ and $\pl_{y_n}-\La^{-n}$. 
This commutativity implies that we may replace $K_+$ everywhere in
(\ref{BC+}) by $K_-$. The reason is that
\[K_-(\pl_{x_n}-\La^n)K_-\inv=K_+(I-F)(\pl_{x_n}-\La^n)K_-\inv
=K_+(\pl_{x_n}-\La^n)(I-F)K_-\inv\]\[=K_+(\pl_{x_n}-\La^n)K_+\inv,\]
and similarly for the definition of $C_n$ given by the second part of (\ref{BC+}). 
(These matrix manipulations are justified using the facts that the rows of $K_+$, the
columns of $K_-\inv$, and their derivatives all belong to $l_2(\bZ)$, and that $F_k$ and its 
derivatives represent bounded operators on $l_2(\bZ_k)$.) Thus we have the second pair of 
relations
\bq \pl_{x_n}-B_n=K_-(\pl_{x_n}-\La^n)K_-\inv,\quad
\pl_{y_n}-C_n=K_-(\pl_{y_n}-\La^{-n})K_-\inv.\label{BC-}\eq
The second statement of (\ref{BCLM}) is obvious once we recognize that the definition of 
$C_n$ may be rewritten
\[C_n={\pl K_+\over\pl y_n}K_+^{-1}+M^n.\]
Similarly the first statement follows from the identity
\[B_n={\pl K_-\over\pl x_n}K_-\inv+L^n,\]
which is equivalent to the first identity of (\ref{BC-}).
To see that the TL equations are satisfied we observe that the operators $\pl_{x_n}-B_n$ 
and $\pl_{y_m}-C_m$ all commute since by (\ref{BC+}) they are simultaneously similar to 
the commuting operators $\pl_{x_n}-\La^n$ and  $\pl_{y_m}-\La^{-m}$.
The commutativity of the $\pl_{x_n}-B_n$ and $\pl_{y_m}-C_m$ is equivalent to the TL 
equations.\sp

\noi{\bf Remark 1}. The definition of $B_n$ in (\ref{BC+}) may be rewritten as
\[B_nK_+={\pl K_+\over\pl x_n}+ K_+\La^n.\]
Since the last term is strictly upper-triangular, we have for the diagonal entries
$B_n(k,k)\,K_+(k,k)=\pl_{x_n}K_+(k,k)$, or 
\[B_n(k,k)=\pl_{x_n}\log\,K_+(k,k).\]
If each $F_k$ represents a trace class operator then Cramer's rule gives the nice formula
\bq K_+(k,k)=\det\,(I-F_{k+1})/\det\,(I-F_k).\label{Kkk}\eq\sp

\noi{\bf Remark 2}. It is familiar how the two-dimensional Toda equations
\bq {\pl^2 q_k\over \pl x\,\pl y}=e^{q_k-q_{k-1}}-e^{q_{k+1}-q_k}\label{xyeq}\eq
lead to one of the equations of the TL hierarchy, 
\bq{\pl B_1\ov\pl y}-{\pl C_1\ov\pl x}+[B_1,\,C_1]=0.\label{TL1}\eq
Here we write $x,\ y$ for $x_1, y_1$ respectively. (See, for example, the introduction to 
\cite{UT}.) In fact we can easily see directly that
\bq q_k(x,y)=\log\,K_+(k,k)\label{qsol}\eq
solves (\ref{xyeq}). From the second relation of (\ref{BC+}) or
(\ref{BCLM}) follows the formula
\[C_1(k+1,k)=K_+(k+1,k+1)/K_+(k,k)=e^{q_{k+1}-q_k},\]
and so taking the $k,k$ entry of (\ref{TL1}) gives
\[{\pl B_1(k,k)\ov\pl y}=C_1(k,k-1)-C_1(k+1,k)=e^{q_k-q_{k-1}}-e^{q_{k+1}-q_k}.\]
Since $B_1(k,k)=\pl q_k/\pl x$ we obtain (\ref{xyeq}).\sp

It remains to write down a class of matrices $F$ satisfying the conditions of the
theorem. If $\mu$ is a measure on $\bC\times\bC$ then
\bq F(i,j)=\int\int\,u^i\,v^j\,e^{\sum [x_n(u^n-v^{-n})+y_n(u^{-n}-v^n)]}d\mu(u,v)
\label{Fij}\eq
satisfies (\ref{F}) as long as the integrands belong to $L_1(\mu)$ for some range of
the parameters $x_n$ and $y_n$. Boundedness of the operators $F_k$ on $l_2(\bZ_k)$  
requires that the support of $\mu$ be contained in the product $\bD \times\bD$ of the 
closed unit discs in \bC. If the support is
contained in the product of the open subdiscs then $F(i,j)$ tends rapidly to 0 as $i$ or
$j$ tends to $+\iy$ and so each $F_k$ is trace class then. We shall not concern ourselves 
here with precise necessary or sufficient conditions. \sp

\setcounter{equation}{0}\renewcommand{\theequation}{2.\arabic{equation}}
\noi{\bf II. Operator solutions to the Toda hierarchy}\sp

In this section and the next we obtain operator solutions to the TL hierarchy from
the matrix solutions above by using two facts about operators. The
first, which is very easy, is that if $A$ and $B$ are operators such that $I-AB$
is invertible then so is $I-BA$ and 
\bq (I-AB)^{-1}=I+A(I-BA)^{-1}B.\label{ABinv}\eq
(This is obvious if the inverses are given by the Neumann series.) The second, which 
is not as easy, is that if $A$ and $B$ are both Hilbert-Schmidt operators, or one operator
is trace class and the other merely bounded, then
\bq \det\,(I-AB)=\det\,(I-BA).\label{ABdet}\eq 
(See, for example, \cite{GK}, sec. IV.1.) It is important that $A$ and $B$ are not required 
to act on the same Hilbert space; $A$ may take a space $H_1$ to a space $H_2$ and 
$B$ take $H_2$ to $H_1$, so that $AB$ acts on $H_2$ and $BA$ acts on $H_1$.

When we say that a particular matrix $K_+$ ``gives a solution to the TL
hierarchy'' we shall mean that the matrices $B_n$ and $C_n$ defined in terms of $K_+$ by 
(\ref{BC+}) are of the form (\ref{BCform}) and satisfy the TL equations. We shall always 
assume that only finitely many of the parameters $x_n,\ y_n$ occur and that the range of
these parameters is such that the exponentials which appear,
such as in (\ref{Fij}), are bounded in the support of the 
corresponding measure.  Throughout, we shall use the notation
\bq E(u,v):=e^{\sum [x_n(u^n-v^{-n})+y_n(u^{-n}-v^n)]/2}.\label{Euv}\eq

The simplest special case,
where the measure $\mu$ in (\ref{Fij}) is supported on the set $u=v$, leads to our first
family of operator solutions to the TL hierarchy. We define
\bq E_k(u):=u^k\,E(u,u)=u^k\,e^{\sum (x_n-y_n)(u^n-u^{-n})/2}.\label{Ek}\eq
\sp

\noi{\bf Theorem 2}. Let $\nu$ be a measure on the unit disc $\bD$ satisfying
\[\int_{\bD}{|E_k(u)|^2\over 1-|u|}\,d\nu(u)<\iy.\]
for all $k\in\bZ$. Then $G_k$, the integral operator on $L_2(\nu)$ with kernel
\bq G_k(u,v)={E_k(u)\,E_k(v)\over 1-uv},\label{Ekkern}\eq
is trace class, and if 1 is not an eigenvalue of any $G_k$ then
\bq K_+(k,j)=\dl_{k,j}+\Big((I-G_k)^{-1}E_j,E_k\Big)\label{KijG}\eq
gives a solution of the TL hierarchy. (The parentheses in the displayed formula denote 
the inner product in $L_2(\nu)$.) The diagonal entries of $K_+$ are also given by
\bq K_+(k,k)=\det\,(I-G_{k+1})/\det\,(I-G_k).\label{KG}\eq\sp

\noi{\bf Proof}. We take the special case of (\ref{Fij}) where $\mu$ is supported on the 
set $v=u$ and is given by $d\mu(u,v)=\dl(u-v)\,d\nu(u)$. Then (\ref{Fij}) becomes
\[F(i,j)=\int\,u^{i+j}\,E(u,\,u)^2\,d\nu(u)\]
and $F_k$ may be written as the product $AB$ where $A$ is the operator from $L_2(\nu)$ to
$l_2(\bZ_k)$ with ``kernel'' $A(i,u)=E_i(u)$
and $B$ is the operator from $l_2(\bZ_k)$ to $L_2(\nu)$ with  kernel $B(u,i)$ given by 
exactly the same formula. They are both Hilbert-Schmidt,
\[\int_{\bD}\,\sum_{i=k}^{\iy}|A(i,u)|^2\,d\nu(u)=\int_{\bD}\,\sum_{i=k}^{\iy}|B(u,i)|^2\,
d\nu(u)=\int_{\bD}{|E_k(u)|^2\over 1-|u|^2}\,d\nu(u)<\iy,\]
and $BA$ is precisely the operator $G_k$. Since $F_k=AB$ is trace class we may apply 
Theorem 1 to conclude that the matrix $K_+$ defined by (\ref{Kkj}) gives a solution to the 
TL hierarchy and (\ref{Kkk}) holds. Applying (\ref{ABinv}) in our situation gives 
(\ref{KijG}) and applying (\ref{ABdet}) gives (\ref{KG}).\sp

\noi{\bf Remark 1}. If $d\nu(u)=p(u)^2\,du$ on the interval $(0,1)\subset \bR$ and
\[\int_0^1 {|E_k(u)|^2\,p(u)^2\over 1-u}\,du<\iy\]
for all $k\in\bZ$ then the conclusion of
the theorem holds if the quantities $E_k(u)$ are replaced
by $E_k(u)\,p(u)$ and the operators act on $L_2(0,1)$. This follows by using the unitary
equivalence between $L_2(\nu)$ and $L_2(0,1)$ given by $f\leftrightarrow pf$. At the other 
extreme, if $\nu$ consists of $N$ masses $p_i^2$ at points $u_i$ then $G_k$ becomes
an $N\times N$ matrix, the Fredholm
determinants become finite determinants and we obtain exponential solutions through the 
formula (recall (\ref{Ek}))
\bq\det\,(I-G_k)=\det\Big(\dl_{i,j}-p_i\,p_j\,{(u_i\,u_j)^k\ov 1-u_i\,u_j}\,
E_0(u_i)\,E_0(u_j)\Big)_{i,j=1,\cdots,N}.\label{sol}\eq\sp

\noi{\bf Remark 2}. If we think of $x_n-y_n$ as a new variable $t_n$ then we obtain 
solutions of the 1-dimensional Toda hierarchy
\[\pl_{t_n}(B_n+C_n)=[B_n,\,C_n].\]
Writing $t=t_1,\,x=x_1,\,y=y_1$ (ignoring the other parameters), keeping in mind that 
$\pl^2/\pl x\,\pl y=-d^2/dt^2$ and substituting into (\ref{xyeq}) we find that  
\[q_k(t)=\log\,K_+(-k,-k)\]
solves the one-dimensional Toda equations
\[{d^2 q_k\over dt^2}=e^{q_{k-1}-q_k}-e^{q_k-q_{k+1}}.\]
In the case of a finite discrete measure we have (\ref{sol}), which becomes
\[\det\,(I-G_k)=\det\Big(\dl_{i,j}-p_i\,p_j\,{(u_i\,u_j)^k\ov 1-u_i\,u_j}\,
e^{t\,(u_i-u_i\inv+u_j-u_j\inv)/2}\Big)_{i,j=1,\cdots,N}.\]
This gives the $N$-soliton solutions of Toda the equations. (See \cite{T}, sec. 3.6.) \sp

\noi{\bf Remark 3}. The kernels (\ref{Ekkern}) and those described in Remark 1 admit
significant generalizations. They are obtained by thinking of 
$\bC\times\bC$ as the union of complex lines $v=\om u\ (\om\in\bC)$ and taking
$d\mu(u,v)=p(\om,u)^2\,d\rho(\om)\,d\nu(u)$, so that (\ref{Fij}) becomes (recall (\ref{Euv}))
\[ F(i,j)=\int\int\,u^{i+j}\,\om^j\,p(\om,u)^2\,E(u,\om u)^2\,d\rho(\om)\,d\nu(u).\]
In the kernels above $\rho$ was a unit mass at $\om=1$.
Now $F_k$ may be written as the product $AB$ where $A$ and $B$ are the operators
from $L_2(\rho\times\nu)$ to $l_2(\bZ_k)$ and from $l_2(\bZ_k)$ to $L_2(\rho\times\nu)$,
respectively, with kernels
\[A(i;\,\om,u)=u^i\,p(\om,u)\,E(u,\,\om u),\quad
B(\om,u;\,i)=(\om u)^i\,p(\om,u)\,E(u,\,\om u).\]
Appropriate assumptions on the measures $d\rho(\om)$ and $d\nu(u)$ and the function
$p(\om,u)$ guarantee that these
are Hilbert-Schmidt operators between $l_2(\bZ_k)$ and $L_2(\rho\times\nu)$. 
The operator $G_k=BA$ on $L_2(\rho\times\nu)$ has kernel
\bq G_k(\om,u;\,\om',v)={(\om uv)^k\ov 1-\om uv}\,p(\om,u)\,p(\om',v)\,
E(u,\,\om u)\,E(v,\,\om' v).\label{OpG}\eq
If we now define
\[E_k(\om,u):=u^k\,E(u,\,\om u)\] 
then we obtain a solution of the TL hierarchy in which (\ref{KijG}) is replaced by 
\[ K_+(k,j)=\dl_{k,j}+\Big((I-G_k)^{-1}\,\om^jpE_j,\,pE_k\Big).\]
The inner product is now taken in $L_2(\rho\times\nu)$. Of course (\ref{KG}) remains the 
same.

If the measure $\rho$ consists of a finite set $\Om$ of unit point masses then we may think 
of $G_k$ as acting
on vector-valued $L_2(\nu)$, the components being indexed by $\om\in\Om$.  The right
side of (\ref{OpG}) gives the $\om,\om'$ entry of the matrix kernel of $G_k$.

Actually, the kernel of $G_k$ can always be transformed to an equivalent scalar kernel by
another $AB\ra BA$ operation. The details of this will be given for another class of 
kernels at the end of the next section.\sp

\setcounter{equation}{0}\renewcommand{\theequation}{3.\arabic{equation}}
\noi{\bf III. More operator solutions to the Toda hierarchy}\sp

The kernels described in the last remark were given
not so much for their inherent interest but because our solutions to the periodic
TL hierarchy and the cylindrical Toda equations arise in a similar way. Now we think of 
$\bC\times\bC$ as the union of complex hyperbolas $v=\om/u$. This time we take 
$d\mu(u,v)=u\inv\,p(\om,u)^2\,d\rho(\om)\,d\nu(u)$ 
(the reason for the factor $u\inv$ in the measure will soon
become apparent) so that (\ref{Fij}) becomes
\[F(i,j)=\int\int\,u^{i-j-1}\,\om^j\,p(\om,u)^2\,E(u,\,\om/u)^2\,d\rho(\om)\,d\nu(u).\]
In this case for the representation of $F_k$ we define
\[ A_k(i;\,\om,u)=u^{i-k}\,p(\om,u)\,E(u,\,\om/u),\quad
B_k(\om,u;\,i)=\om^i\, u^{k-i-1}\,p(\om,u)\,E(u,\,\om/u).\]
The situation is more awkward now, and we begin with the assumption 
that $d\nu(u)$ and $d\rho(\om)$ are finite measures supported on annuli
\[r_1\leq |u|\leq r_2<1,\quad 0<s_1\leq |\om|\leq s_2\]
where $s_2<r_1$. We call this ``the annulus condition''. We also assume that the function
$p(\om,u)\,E(u,\,\om u)$ belongs to $L_2(\rho\times \nu)$. It is easy to see that 
both operators $A_k$ and $B_k$ are then Hilbert-Schmidt, $F_k=A_kB_k$ and the operator 
$G_k=B_kA_k$ on $L_2(\rho\times \nu)$ has kernel
\bq G_k(\om,u;\,\om',v)={\om^k\ov u-\om v}\,p(\om,u)\,E(u,\,\om/u)\,
p(\om',v)\,E(v,\,\om'/v).\label{Gk}\eq
If we define now
\[E_k(\om,u):=u^k\,E(u,\,\om/u)\]
then the solution to the TL hierarchy is given by the formula
\bq K_+(k,j)=\dl_{k,j}+\Big((I-G_k)^{-1}\,\om^jpE_{k-j-1},\,pE_0\Big),\label{KGk}\eq
and (\ref{KG}) holds as usual.  

The very restrictive condition imposed on $d\nu(u)$ and $d\rho(\om)$ makes this 
not very interesting. But the formulas make sense much more generally and we might expect 
them to give solutions whenever the operators $G_k$ with kernel defined by (\ref{Gk}) are 
trace class and the $I-G_k$ are invertible. Although we cannot prove such a general result 
we can prove enough for our purposes. We denote the supports of the finite measures 
$d\nu(u)$ and $d\rho(\om)$ by $U$ and $\Om$ respectively and use the familiar notation
\[UU\inv:=\{uv\inv:u, v\in U\}.\]
The annulus condition implies that $UU\inv$ is also contained in
an annulus and that $\Om$ is inside the inner boundary of this annulus. 
In particular $\Om$ can be shrunk down to 0 without in the process intersecting $UU\inv$. As
we shall now show, a quasi-topological condition like this on $U$ and $\Om$ is all that
we need. Note that just the assumption $\Om\cap UU\inv=\emptyset$ implies that $G_k$ has 
$C^{\iy}$ kernel and so is trace class.\sp

\noi{\bf Theorem 3}. Assume that $pE_0\in L_2(\rho\times \nu)$ and that the supports $U$ and 
$\Om$ are compact subsets of $\bC\backslash0$ with the property that there is a path from 
0 to 1 in $\bC$ such that $\be\Om\cap UU\inv=\emptyset$
for all $\be$ in the path. Assume also that each $I-G_k$ is invertible. Then 
(\ref{KGk}) gives a solution of the TL hierarchy and (\ref{KG}) holds.\sp

\noi{\bf Remark}. For the proof it will be convenient to widen the meaning of the
phrase ``$K_+$ gives a solution to the TL hierarchy'' to the case where not all
entries of $K_+$ are defined. The TL hierarchy consists of an infinite
number of scalar equations each of which depends on only finitely many entries of
the matrices $B_n$ and $C_n$, and these in turn are determined from (\ref{BC+}) using
only finitely many entries of $K_+$. If $K_+$ is only a partially defined matrix
then our phrase will mean that all those scalar equations of the TL hierarchy which
make sense are satisfied. This will be the case if in the statement of the
theorem not all $I-G_k$ are invertible. This occurs for those
values of the paramenters $x_n$ and $y_n$ for which $\det\,(I-G_k)=0$ and at these values
certain entries of the solution matrices become singular. Even if all $I-G_k$ are invertible 
this concept, in proving the theorem, will allow us to deal with finitely many $k$ at a 
time rather than all $k$ simultaneously.\sp

\noi{\bf Proof}. For nonzero complex numbers $\al$ and $\be$ let $\nu_{\al}$ and 
$\rho_{\be}$ be the measures on $\al U$ and $\be\Om$ defined by
\[\nu_{\al}(V)=\nu(\al\inv V),\quad \rho_{\be}(W)=\rho(\be\inv W).\]
If we first choose $\al$ small enough and then choose $\be$ small enough these
measures will satisfy the annulus condition. Thus if we replace $\nu$ by $\nu_{\al}$,
$\rho$ by $\rho_{\be}$ and $p(\om,u)$ by $p(\be\inv \om,\al\inv u)$ then the corresponding 
operators give a solution to the TL hierarchy by the corresponding formula (\ref{KGk}), and
(\ref{KG}) holds. If 1 
is an eigenvalue of some of these operators then we obtain a partial solution, as described 
in the remark. These operators act on $L_2(\rho_{\be}\times \nu_{\al})$ but by means of 
the variable changes $\om\ra\be\om, u\ra\al u$ we obtain operators $G_k^{(\al,\be)}$ on 
$L_2(\rho\times \nu)$ which give a 
solution by the corresponding formulas (\ref{KGk}). Their kernels are 
\[{\al\inv\,\om^k\ov u-\be\om v}\,p(\om,u)\,
E(\al u,\,\be \om/\al u)\,p(\om',v)\,E(\al v,\,\be \om'/\al v).\]
These will give a solution to the TL hierarchy if $\al$ lies in some annulus
$r_1<|\al|<r_2$ and $\be$ in some punctured disc $0<|\be|<s$.

Consider a finite set $k_1,\cdots,k_n$ of $k$ for each of which $I-G_k$ is invertible for
particular values of the parameters $x_n$ and $y_n$.
We are going to show that the TL equations which are defined 
using the formula (\ref{KGk}) for these $k$ are correct for these values of the parameters. 
Choose any $\be$ with $0<|\be|<s$. 
A glance at the form of its kernel shows that the operator
$G_k^{(\al,\be)}$ is well-defined for all $\al\in\bC\backslash0$ and is analytic in $\al$.
The set of $\al$ such that $I-G_k^{(\al,\be)}$ is not invertible for one of our $k$
is the union of the set of zeros of the functions (of $\al$) $\det\,(I-G_k^{(\al,\be)})$. 
This is a discrete subset of $\bC\backslash0$ varying continuously with the parameters. 
It follows that there is a path running from
a point in $r_1<|\al|<r_2$ to $\al=1$ such that everywhere on a neighborhood of this
path, and on neighborhoods of our given parameter values, all operators $I-G_k^{(\al,\be)}$ 
are invertible. The matrix entries which arise in
those TL equations which are defined using only $k_1,\cdots,k_n$ are analytic functions
of $\al$ and since these equations are satisfied for $r_1<|\al|<r_2$ they must persist
by analytic continuation to a neighborhood of $\al=1$. In other words, the operators
$G_k^{(1,\be)}$ give a partial solution to the TL hierarchy for $k=k_1,\cdots,k_n$
whenever $|\be|<s$.

Now we can apply the same argument to $\be$. Our hypothesis implies that there is
a path running from a point in $0<|\be|<s$ to $\be=1$ on a neighborhood of which 
$G_k^{(1,\be)}$ is an analytic family of trace class operators: the denominator in the 
formula for the kernel will remain nonzero. By deforming the path slightly if 
necessary we can assure that for $\be$ on the path all operators $I-G_k^{(1,\be)}$
are invertible. Analytic continuation as before now shows that the $G_k^{(1,1)}$, in other
words our given operators $G_k$, give a solution to the TL hierarchy.\sp

A limiting argument applied to a special case of the above will give a class of solutions
to the TL hierarchy which include periodic ones. The measure $\nu$ will now be supported
on $\bR^+$, the nonnegative real numbers in \bC. Thus we shall have a case where the support
of $\nu$ is neither compact nor contained in $\bC\backslash0$. This
is the reason a limiting argument is necessary. \sp

\noi{\bf Theorem 4}. Assume $\Om$ is a compact subset of $\bC\backslash\bR^+$, the measure 
$\nu$ is supported on $\bR^+$
and $pE_i\in L_2(\rho\times \nu)$ for all $i\leq0$. Then (\ref{KGk}) 
gives a solution of the TL hierarchy and (\ref{KG}) holds.\sp

\noi{\bf Remark}. Our basic assumtion of boundedness of $E(u,\,\om/u)$ will imply that it 
vanishes exponentially at $u=0$ and $u=\iy$. It follows that the function $p(\om,u)$ can be 
very general.\sp

\noi{\bf Proof}. For $r>1$ let $P_r$ denote multiplication by the characteristic function
of the interval $[r\inv,\,r]$ and define $G_k^{(r)}:=P_r\,G_k\,P_r$. These may be thought of 
alternatively as the operators $G_k$ which arise when $\nu$ is replaced by its restriction 
to $[r\inv,\,r]$. The assumption of Theorem 3 is satisfied for this measure: we can take 
our path from $\be=0$ to $\be=1$ to be the line segment. Thus for each $r$ the corresponding
operators $G_k^{(r)}$ and formulas (\ref{KGk}) give solutions (or partial 
solutions) to the TL hierarchy, and (\ref{KG}) holds. 

We shall show that the trace norm of the operator $G_k-G_k^{(r)}$ tends to 0 as $r\ra\iy$.
We write $G_k-G_k^{(r)}=(I-P_r)\,G_k+P_r\,G_k\,(I-P_r)$ and apply Lemma 1 of the
appendix to each of the operators on the right. Denote by $\ch_r(u)$ the characteristic 
function of the $u$-interval $[r\inv,\,r]$, and recall (\ref{Gk}). In the first application 
of the lemma $q_1=(1-\ch_r)\,\om^kpE_0,\ q_2=pE_0$ and 
in the second application $q_1=\ch_r\,\om^kpE_0,\ q_2=(1-\ch_r)\,pE_0$. In both cases one 
of the integrals in the statement of the lemma will tend to 0 as $r\ra\iy$. Notice that
all we need here is that  $pE_{-1/2}\in L_2(\rho\times \nu)$. 

All functions which appear in the inner products in (\ref{KGk}) are of the form 
$\om^jpE_i$ for $i\leq 0$, and these belong to $L_2(\rho\times \nu)$ by our main assumption.
It follows that the functions of the parameters $x_n$ and $y_n$
which appear the TL equations obtained from
the $G_k$ using (\ref{KGk}) are the limits of the corresponding functions obtained from
the $G_k^{(r)}$. The same is true of the derivatives of these functions. (This is
automatic because the functions are locally bounded and analytic in the parameters.)
Since the equations are satisfied for each $r$ the equations must be satisfied by the 
limiting functions. Trace norm convergence of the operators is needed
to deduce (\ref{KG}): we know that it holds for the solution corresponding to $G_k^{(r)}$
and so
\[K_+(k,k)=\lim_{r\ra\iy}{\det\,(I-G_{k+1}^{(r)})\ov\det\,(I-G_k^{(r)})}
={\det\,(I-G_{k+1})\ov\det\,(I-G_k)}.\]
The convergence of the determinants requires trace norm convergence of the operators.\sp

We are now going to transform (\ref{Gk}) into a kernel on $L_2(\nu)$  by another 
$AB\ra BA$ operation. (We think of this as a scalar-valued kernel on $L_2(\nu)$ whereas (\ref{Gk}) 
could be thought of as an $L_2(\rho)$ operator-valued kernel or, in case $\Om$ is finite,
a matrix-valued kernel.) Now $B$ will be the operator from 
$L_2(\rho\times \nu)$ to $L_2(\nu)$ which is multiplication by $pE_0$ followed by 
integration with respect to $d\rho$ over $\Om$. If 
$pE_0$ is bounded then this will be a bounded operator. The operator $A$ from
$L_2(\nu)$ to $L_2(\rho\times \nu)$ has kernel $\om^k\,p(\om,u)\,E(u,\,\om/u)\,/(u-\om v)$.
(In other words one takes a function $g(v)$, multiplies by this kernel, and then
integrates with respect to $d\nu(v)$.)
The operator $\tl G_k=BA$ on $L_2(\nu)$ has kernel
\bq\tl G_k(u,v)=\int_{\Om}\om^k{p(\om,u)^2\,E(u,\,\om/u)^2\ov u-\om v}\,d\rho(\om).
\label{Gkt}\eq
In order to apply (\ref{ABinv}) we have to know that $A$ is also bounded, and in order
to apply (\ref{ABdet}) we have to know that it is even trace class, since $B$ is surely not
Hilbert-Schmidt. Fortunately, we have Lemma 2 of the appendix, and so we can immediately
give a variant of Theorem 4 for this scalar kernel. The condition imposed on $\nu$ is
a little awkward but it is satisfied by Lebesgue measure, the most interesting case. We 
shall not write down the analogue of
(\ref{KGk}) here since it may be obtained from (\ref{KGk}) itself by applying (\ref{ABinv}).
\sp

\noi{\bf Theorem 4$'$}. In addition to the assumptions of Theorem 4, assume that $pE_0$ is 
bounded and that for some $\dl\in(0,2)$
\[\int_0^{\iy}{d\nu(u)\ov u^{2-\dl}+1}<\iy\]
and $u^{-1-\dl/2}p^2E_0^2\in L_2(\rho\times \nu)$. Then the 
operators $\tl G_k$ give a solution of the TL hierarchy if all $I-\tl G_k$ are 
invertible, and  we have
\bq K_+(k,k)=\det\,(I-\tl G_{k+1})/\det\,(I-\tl G_k).\label{KGt}\eq\sp

We now show that the special case of these kernels where $\nu$ is Lebesgue measure and
$p$ is independent of $u$ (in which case it can be incorporated into the measure $\rho$) 
gives solutions to the cylindrical Toda equations
\[{d^2 q_k\over dt^2}+t\inv\,{dq_k\over dt}=e^{q_k-q_{k-1}}-e^{q_{k+1}-q_k}.\]
We consider only the scalar kernel $\tl G_k$ although the case of $G_k$ is no
different. Again we ignore all parameters except $x=x_1$ and $y=y_1$, so (\ref{Gkt})
becomes
\[\tl G_k(u,v)=\int_{\Om}\om^k\,{E(u,\,\om/u)^2\ov u-\om v}\,d\rho(\om)\]
with
\[E(u,\,\om/u)^2=e^{x(1-\om^{-1})u+y(1-\om)u\inv}.\]
If we make the variable change $u\ra\sqrt{y/x}\,u$ then the new kernels, which have the
same determinants, are
\[\int_{\Om}\om^k\,{e^{\sqrt{xy}\,[(1-\om^{-1})u+(1-\om)u\inv]}\ov u-\om v}
\,d\rho(\om).\]
(This is where we use the fact that $\nu$ is Lebesgue measure and $p$ is independent
of $u$.) Thus the determinants 
in (\ref{KGt}) are functions of $xy$, and if we set $t=2\sqrt{xy}$
then the two-dimensional Toda equations (\ref{xyeq}) become the cylindrical Toda 
equations. Solutions are given by (\ref{qsol}) and (\ref{KGt}).\sp

\setcounter{equation}{0}\renewcommand{\theequation}{4.\arabic{equation}}
\noi{\bf IV. The} {\boldmath $l$}{\bf -periodic Toda hierarchy}\sp

A doubly-infinite matrix $M(i,j)$ is called $l$-periodic if $M(i+l,j+l)=M(i,j)$ for
all $i,j$. The solutions $B_n,\ C_n$ of the TL hierarchy given by (\ref{BC+}) are 
$l$-periodic if the matrix $K_+$ is. And the matrix $K_+$ given by (\ref{KGk}) is 
$l$-periodic if $\Om$ is contained in the set of $l$th roots of unity, as is easily seen
by referring to (\ref{Gk}) and (\ref{KGk}). Of course since $\Om$ is disjoint from $\bR^+$
the root 1 must be omitted, so we may take $\Om$ to be the set of $l$th roots of 
unity other than 1. Assuming $\rho(\{\om\})=1$ for each of these roots the formula (\ref{Gk}) 
for the kernel, now a matrix kernel, may be written 
\bq G_k(u,v)=\Big({\om^k\ov u-\om v}\,p_{\om}(u)\,E(u,\,\om/u)\,
p_{\om'}(v)\,E(v,\,\om'/v)\Big)_{\om,\om'\in\Om}\label{lGk}\eq
and the hypothesis of Theorem 4 becomes 
\[\int_0^{\iy} |p_{\om}(u)\,u^iE(u,\,\om/u)|^2\,d\nu(u)<\iy\] 
for all $\om$ and $i\leq0$. The scalar kernel of $\tl G_k$ becomes
\bq\tl G_k(u,v)=\sum_{\om}\om^k\,{p_{\om}(u)^2\,E(u,\,\om/u)^2\ov u-\om v},\label{Gsum}\eq
with additional assumptions coming from the statement of Theorem 4$'$. 

The case where only one $\om$ occurs is especially simple. If $G$ is the operator with
either kernel
\[G(u,v)={p(u)\,E(u,\,\om/u)\,p(v)\,E(v,\,\om/v)\ov u-\om v}\ \mbox{or}\ 
{p(u)^2\,E(u,\om/u)^2\ov u-\om v}\]
then $G_k$ resp. $\tl G_k$ equals $\om^k\,G$ and (\ref{KG}) becomes
\bq K_+(k,k)=\det\,(I-\om^{k+1}\,G)/\det\,(I-\om^k\,G).\label{perG}\eq
Of course the assumptions are slightly different in the two cases. If $\nu$ consists of
$N$ masses $p_i^2$ at the points $u_i$ then the Fredholm determinants become
finite determinants,
\[\det\,(I-\om^k\,G_k)=\det\Big(\dl_{i,j}-\om^k\,p_i\,p_j\,{E(u_i,\,\om/u_i)\,
E(u_j,\,\om/u_j)\,\ov u_i-\om\,u_j}\,\Big)_{i,j=1,\cdots,N}.\]
\sp

\noi{\bf Remark 1}. Observe that for $l=2$ only $\om=-1$
occurs and so each term in (\ref{perG}) is the ratio of the determinants 
$\det\,(I\pm G)$. \sp

\noi{\bf Remark 2}. If in (\ref{lGk}) or (\ref{Gsum}) $p_{\om}(u)$ is independent of $u$ 
then we are in the case considered at the end of Section III. Therefore if $\nu$ is
Lebesgue measure and we set $t=2\sqrt{x_1y_1}$ we
obtain periodic solutions of the cylindrical Toda equations through the formulas 
(\ref{qsol}) and (\ref{KG}) or (\ref{KGt}).\sp 

\setcounter{equation}{0}\renewcommand{\theequation}{5.\arabic{equation}}
\noi{\bf V. Appendix}.\sp

We prove here the lemmas needed for Theorems 4 and 4$'$. First we shall prove a sublemma, 
which gives a family of estimates for the trace norm of the operator from 
$L_2(\rho'\times \nu)$ to $L_2(\rho\times \nu)$
with kernel 
\bq{q_1(\om,u)\,q_2(\om',v)\ov u-\om v}.\label{opkern}\eq
There will be an inequality for each positive funtion $\ph(s)$ defined on $\bR^+$. 
We denote the Laplace transform of $\ph(s)$ by $\Phi$ and the Laplace transform of 
$\ph(s)\inv$ by $\Psi$. \sp

\noi{\bf Sublemma}. Let $\rho$ be a finite measure supported on a compact set 
$\Om\subset\bC\backslash\bR^+$ and $\nu$ a measure supported on $\bR^+$. (The measure $\rho'$
is arbitrary.) Then there is a
constant $m$ depending only on $\Om$ such that the square of
the trace norm of the operator with kernel (\ref{opkern}) is at most $m^{-2}$ times
the square root of
\[\int_0^{\iy}\int_{\Om}|q_1(\om,u)|^2\,\Phi(mu)\,d\rho(\om)\,d\nu(u)
\cdot\int_0^{\iy}\int_{\Om}\,|q_2(\om',u)|^2\,\Psi(mu)\,d\rho'(\om')\,d\nu(u).\]\sp

\noi{\bf Proof}. Write $-\om=r^2\,e^{2i\th}$. Then $r$ is bounded and bounded
away from 0 for $\om\in\Om$ and we may take $|\th|\leq{\pi\ov2}-\dl$ for some $\dl>0$. 
With this notation we write
\[{1\ov u-\om v}={r^{-1}e^{-i\th}\ov r^{-1}e^{-i\th}u+re^{i\th}v},\]
which has the integral representation
\[\int_0^{\iy}r^{-1}e^{-i\th}\,e^{-sr^{-1}\,e^{-i\th}u}\,e^{-sre^{i\th}v}\,ds.\]
It follows that the kernel of our operator has the integral representation
\bq\int_0^{\iy}r^{-1}e^{-i\th}q_1(\om,u)\,e^{-sr^{-1}\,e^{-i\th}u}\,
q_2(\om',v)\,e^{-sre^{i\th}v}\,ds.\label{T}\eq
From this representation it follows that for any choice of $\ph(s)$ we can factor the
operator 
as the product $AB$ where $A$ is the integral operator from $L_2(\bR^+)$ (with Lebesgue
measure) to $L_2(\rho\times \nu)$ with kernel
\[A(\om,u;\;s)=r^{-1}e^{-i\th}q_1(\om,u)\,\ph(s)^{1/2}\,e^{-sr^{-1}\,e^{-i\th}u}\]
and $B$ is the integral operator from $L_2(\rho'\times \nu)$ to $L_2(\bR^+)$ with kernel
\[B(s;\;\om',u)=q_2(\om',u)\,\ph(s)^{-1/2}\,e^{-sre^{i\th}u}.\]
Let $m>0$ be any constant less than or equal to $r\cos\th$ and to $r\inv\cos\th$ for all 
$\om\in\Om$. (Notice that $\cos\th\geq\sin\dl$ for all 
$\om\in\Om$.) Then the square of the Hilbert-Schmidt norm of $A$ is at most
$m^{-2}$ times
\[\int\int\int_0^{\iy}|q_1(\om,u)|^2\,\ph(s)\,e^{-smu}\,ds\,d\rho(\om)\,d\nu(u)=
\int\int|q_1(\om,u)|^2\,\Phi(mu)\,d\rho(\om)\,d\nu(u).\]
Similarly the square of the Hilbert-Schmidt norm of $B$ is at most
\[\int\int|q_2(\om',u)|^2\,\Psi(mu)\,d\rho'(\om')\,d\nu(u),\]
which establishes the sublemma.\sp

\noi{\bf Lemma 1}. The trace norm of the integral operator on $L_2(\rho\times \nu)$
with kernel (\ref{opkern}) is at most a constant depending only on $\Om$ times the
square root of
\[\int_0^{\iy}\int_{\Om}u\inv\,|q_1(\om,u)|^2\,d\rho(\om)\,d\nu(u)
\cdot\int_0^{\iy}\int_{\Om}\,u\inv\,|q_2(\om,u)|^2\,d\rho(\om)\,d\nu(u).\]\sp

\noi{\bf Proof}. In the sublemma take $\rho'=\rho$ and $\ph(s)\equiv 1$.\sp

\noi{\bf Lemma 2}. For any $\dl\in(0,2)$ the trace norm of the integral operator from 
$L_2(\nu)$ to $L_2(\rho\times \nu)$ with kernel $q(\om,u)/(u-\om v)$
is at most a constant depending only on $\Om$ and $\dl$ times the square root of
\[\int_0^{\iy}\int_{\Om}(u^{-2-\dl}+1)\,|q(\om,u)|^2\,d\rho(\om)\,d\nu(u)
\cdot\int_0^{\iy}{d\nu(u)\ov u^{2-\dl}+1}.\]\sp

\noi{\bf Proof}. In the sublamma we take $\rho'$ to be a unit point mass, so that
$L_2(\rho'\times \nu)$ may be identified in the obvious way with $L_2(\nu)$. We 
take $\ph(s)=s^{-1+\dl}$ for $s\leq1$ and $\ph(s)=s^{1+\dl}$ for $s\geq1$. Then it is
easy to see that
\[\Phi(s)=\left\{\begin{array}{ll}
O(t^{-2-\dl}) & \mbox{when $t\ra0$}\\ O(1) & \mbox{when $t\ra\iy$,}\end{array}\right.\quad
\Psi(s)=\left\{\begin{array}{ll}
O(1) & \mbox{when $t\ra0$}\\ O(t^{-2+\dl}) & \mbox{when $t\ra\iy$.}\end{array}\right.\]
Combining these estimates with the sublemma gives the statement of the lemma.\sp\sp

\begin{center}{\bf Acknowledgements}\end{center}

The author thanks Fritz Gesztesy, David Sattinger and Pierre Van Moerbeke
for helpful electronic conversations.  
Special thanks go to Craig Tracy, whose ongoing collaboration with the 
author led to this work and whose advice during its preparation was invaluable.  
Research was supported by the National Science Foundation through grant DMS-9424292.
\sp\sp

\end{document}